**Shapeable planar Hall sensor with a stable sensitivity under concave and convex bending**

B. Özer, H. Pişkin, and N. Akdoğan

*Gebze Technical University, Department of Physics, 41400 Gebze, Kocaeli, Turkey*
(Dated: 6 December 2018)

A shapeable planar Hall sensor has been fabricated by lift-off a bilayer structure of NiFe(10 nm)/IrMn (8 nm) grown on a Kapton/PDMS substrate without using a buffer layer. The sensor exhibits a magnetic field sensitivity of 0.74 $\mu$V/Oe.mA and provides a stable response under repetitive ON/OFF experiments. The concave and convex bending measurements indicate that the AMR voltage of the sensor is very sensitive to the stress and strain. However, the planar Hall sensitivity is independent from them with a stable linear region and unchanged peak positions in the magnetic field axis. This type of behaviour has been observed for the first time in the planar Hall sensors. Therefore, this novel device can fulfill simultaneous multifunctional sensing of small magnetic fields, stress and strain. This makes it promising for tactile sensing applications in humanoid robotics.

## I. INTRODUCTION

Shapeable magnetoresistive devices became one of the most important technological research fields of the last years. It is expected that the proposed device should possess at least one of the properties such as flexible, printable, and stretchable. Foreseeable applications of highly sensitive, cost effective and re-shapeable magnetoelectronics also include their integration into the microfluidics, lab-on-skin and lab-on-a-chip platforms.[1–4] Among the magnetoresistive-based chips, planar Hall effect (PHE) sensors have the advantages of high signal-to-noise ratio, small offset voltage, and linear response at small fields. Therefore, there has been extensive research to increase their sensitivity by optimizing sensor materials and sensor architectures.[5–9]

Although the PHE sensors can easily be grown on conventional rigid substrates, there are many issues to successfully fabricate them on shapeable surfaces. Difficulties in substrate preparation, cleaning and lithographic processes, troubles in deposition of ultra-thin sensor layers on rough surfaces, problems during thermal treatments and contact preparation for device characterization, physical weakness of the sensor structures under strong stress and strain, sensor self heating due to the applied current can be mentioned as main challenges.

In the literature, there are few efforts to fabricate PHE sensors on a shapeable surface. Recently, *Oh* and coworkers developed a magnetoresistive sensor grown a PEN substrate with a spin-valve structure in order to detect pathogenic bacteria.[10] However, they did not provide bending experiments to check the sensor performance under stress or strain. They have also fabricated a ring junction consist of a bilayer structure on a PEN substrate and observed a decrease in the sensor sensitivity during bending due to the stress induced anisotropy.[11]

In this work, we present a shapeable planar Hall sensor which exhibits a stable PHE sensitivity and a bending sensitive anisotropic magnetoresistance (AMR) under concave and convex bending conditions. This novel device has been fabricated on a flexible Kapton/PDMS substrate with a Hall-bar geometry of a bilayer structure. We have observed a PHE sensitivity of 0.74 $\mu$V/Oe.mA at room temperature. The real-time experiments reveal that the sensor states are very stable under different magnetic fields and the sensor output voltage goes back to the initial position when the field is zero. While the AMR signal of the produced sensor is very sensitive to the stress and strain during bending experiments, the PHE sensitivity is independent from them. The mechanisms behind the different behaviours of the AMR and PHE signals are also discussed.

## II. SENSOR FABRICATION AND DETAILS OF THE BENDING EXPERIMENTS

The PHE sensor has been fabricated by optical lithography on a Kapton/PDMS substrate. First of all, a commercial Kapton tape has been sticked on a transparent and flexible PDMS prepared by a pre-polymer gel and a liquid curing agent with a ratio of 10:1. A Hall-bar geometry with dimensions of $60\mu m \times 60\mu m$ has been formed by lift-off a photoresist layer coated on the Kapton/PDMS substrate. Then, the sensor structure of NiFe(10)/IrMn(8)/Pt(3)(nm) has been deposited at room temperature by magnetron sputtering with a base pressure of 1.5×10$^{-9}$ mbar. It is worthy of note that the required sample structure has been successfully grown without using a buffer or seed layer. Fig. 1(a) shows a photograpgh of the cleanroom fabricated PHE sensor taken from the backside of the device. A patterned and continuous film parts of the sensor structure can bee seen in the picture. A microscopic image of the Hall-bar structure and the schematic drawing of the measurement geometries are given in Fig. 1(b). The easy axis of the magnetization in the sensing layer has been formed parallel to the current ($i$) axis by growth-induced anisotropy without applying magnetic field during the deposition. Thus, the exchange-bias field ($H_{ex}$) has been also aligned parallel to the current direction. A sensor current of 0.3 mA has been applied using a Keithley 2400 sourcemeter. The AMR and planar Hall voltages have been measured at room temperature by a Keithley 2002 multimeter.

Fig. 1(c) displays the bending measurement setup integrated into the home-made transport measurement system.[12] The setup consists of a flexible, transparent and long ribbon equipped with a printed circuit board (PCB). The shapeable PHE sensor has been mounted on the ribbon, and the connections from the sensor to the PCB have been made by using silver paint and shielded copper wires. The concave and convex bending configurations have been produced by changing the radius of the ribbon, as schematically shown in Fig. 1(d). The actual bending radius of the sensor on the ribbon has been determined from the top-view photograph by using a photo editing program.

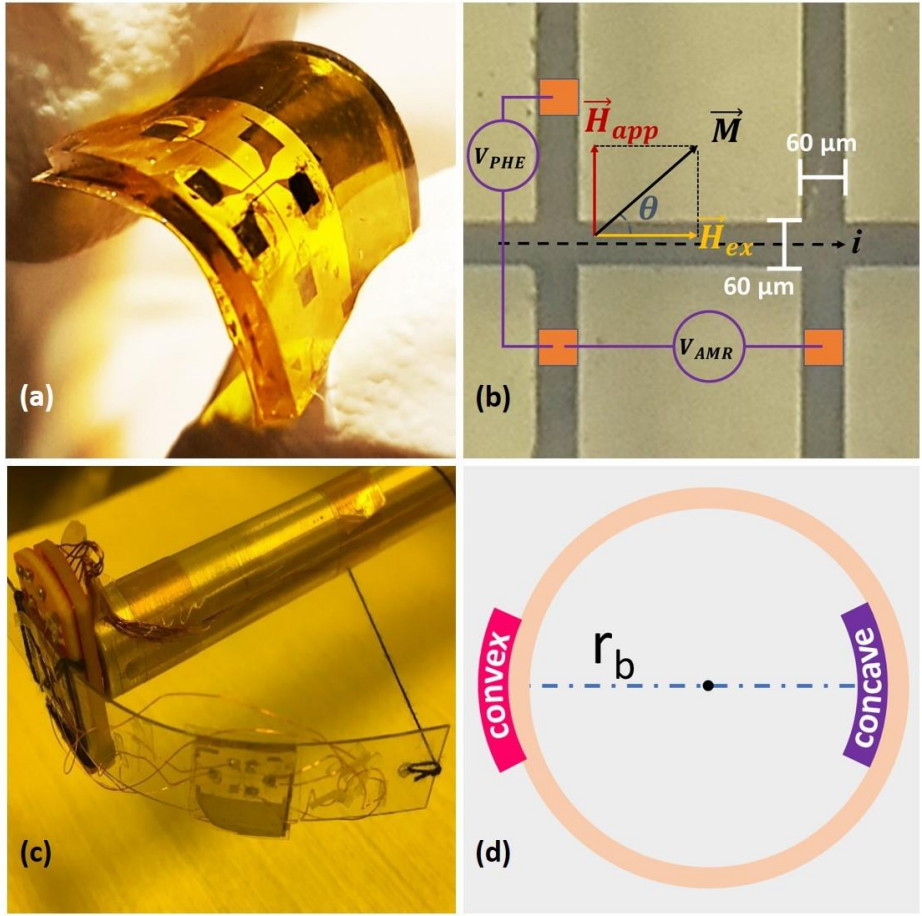

FIG. 1. (a) Photograph of a shapeable PHE sensor on a Kapton/PDMS substrate. (b) Schematic representation of the measurement geometries for PHE and AMR experiments on a microscopic picture of the Hall-bar structure. (c) Picture of the bending measurement setup consists of a ribbon holder and a PCB. Shapeable PHE sensor can be seen on the transparent ribbon holder. (d) A schematic drawing of the sensor on a ribbon holder for concave and convex configurations and representation of the bending radius ($r_b$).

## III. RESULTS AND DISCUSSION

The magnetic and exchange bias properties of the PHE sensor have been investigated by using a magneto-optical Kerr effect (MOKE) setup.[12] Fig. 2(a) presents easy axis L-MOKE hysteresis curve of the bilayer structure taken from the continuous part of the sample at room temperature. We have observed a complete shift of the hysteresis curve to the negative field side with an exchange bias field of 91 Oe. Then, we have carried out PHE experiments by sweeping the in-plane magnetic field perpendicular to the applied current. Fig. 2(b) indicates a clear PHE voltage profile of the shapeable PHE sensor with a linear response for a wide range of the applied field. A hysteretic behaviour has been also observed in the PHE signal. This is attributed to an incoherent magnetization rotation within the sensor structure and discussed in a previous work for a trilayer-based system.[9]

We have also performed real-time experiments in order to check the stability of the sensor states under different magnetic fields and to investigate the reproducibility of the sensor signal. Fig. 2(c) shows PHE voltages recorded at room temperature for 5 minutes as a function of magnetic field varies from zero to 20 Oe with a step of 4 Oe. The data exhibits very stable sensor states at different magnetic fields. The inset in Fig. 2(c) indicates the sensor output voltages measured for successive magnetic fields of 0 Oe and 4 Oe. We have demonstrated that the sensor states are repeatable and sensor signal goes back to the offset value when the applied field is zero.

Furthermore, we have investigated the AMR voltage and AMR ratio as a function of bending radius. Fig. 3(a) depicts AMR signals for concave, flat and convex conditions of the shapeable PHE sensor. Actually, we have recorded AMR signals for each bending radius. But, we have presented here only three of them in order to increase the visibility of the data. It is clearly seen that the AMR signal has been increased when the sensor is bent in convex form. However, it has been decreased in concave condition. In contrast to the AMR voltage, we have observed an increase in the AMR ratio when the amount of the concave bending has been increased. But, it has been decreased under convex bending (Fig. 3(b) and (c)). We have also realized that the AMR ratios for the released conditions slightly differ from the value when the sensor is flat. The difference between the flat and released conditions has been occurred due to the flexible ribbon used in the bending setup, which does not completely turn back to its initial position when it has been released.

The PHE voltage profiles recorded for concave, flat and convex conditions are given in Fig. 4. There is no substantial difference between the PHE voltage profiles measured at different bending conditions. Besides, Fig. 4(b) and (c) indicate that the linear parts of the PHE signals coincide completely with each other. In addition, the maximum and minimum PHE voltages have been observed at the same magnetic fields for both concave and convex conditions. This results indicate that both the PHE voltage and PHE sensitivity of the produced sensor are independent from the bending.

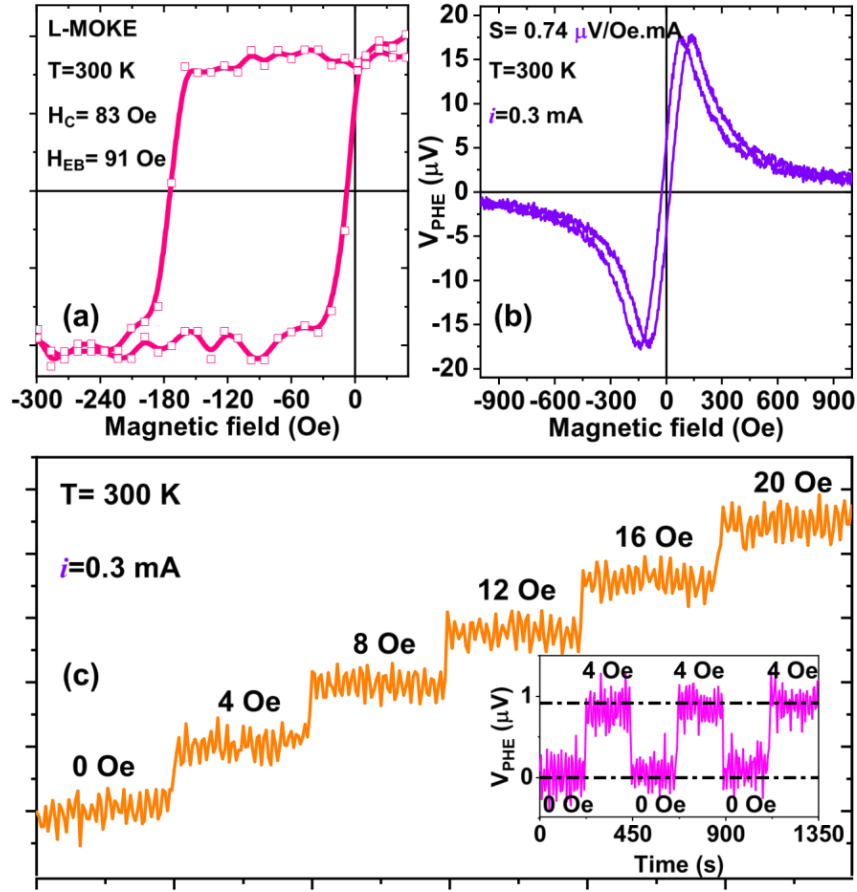

FIG. 2. (a) L-MOKE hysteresis loop of the NiFe/IrMn bilayer structure grown on Kapton/PDMS substrate. (b) The PHE signal profile of the sensor under a cycle of applied magnetic field. (c) Real-time PHE voltage profile of the sensor measured under increasing steps of the magnetic field. Inset shows the sensor output signal as a function of the time for repetitive magnetic fields of 0 Oe and 4 Oe, indicating OFF and ON states, respectively. Solid lines are guide for the eye.

Indeed, both the AMR and PHE data presented in Figs. 3 and 4 are in good agreement with the theory. When the sensor is bent in concave configuration, both the parallel ($\rho_\parallel$) and perpendicular ($\rho_\perp$) resistivities have been decreased due to the stress produced on the sensor. In contrast, they have been increased under the strain for convex condition. Both the

AMR (Eq. 1) and PHE (Eq. 2) voltages depend on the $\rho_\parallel$ and $\rho_\perp$ of the sensing layer.[13] However, their dependence on the resistivities are not same. If the same amount of change occurs in the resistivities under the stress and strain, the change in the resistivity difference

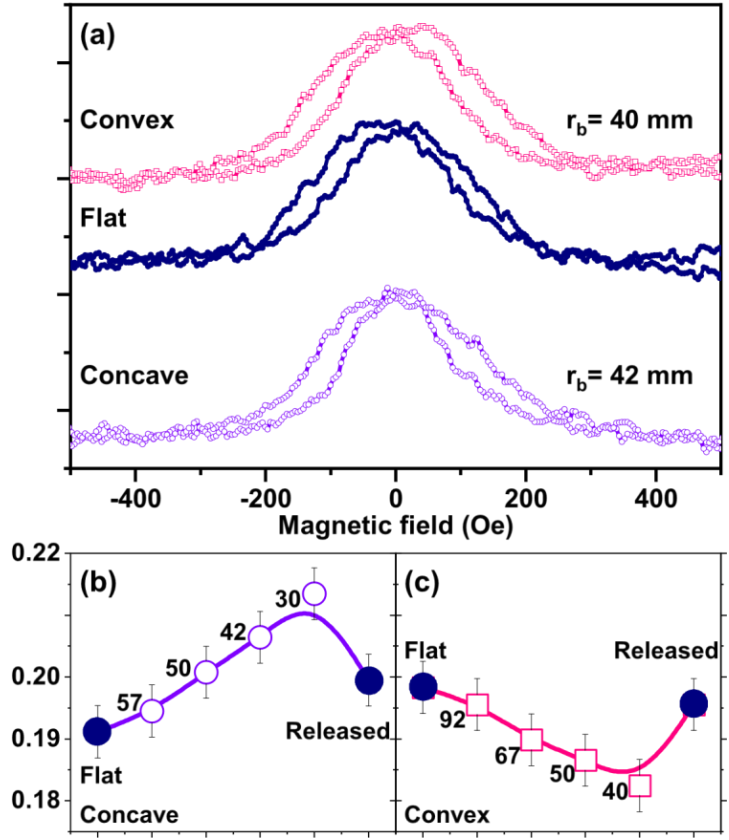

FIG. 3. (a) The AMR signals of the sensor taken for convex, flat and concave conditions. $r_b$ denotes the bending radius. (b) and (c) present the AMR ratio and error bars as a function of bending radius for concave (open circles) and convex (open squares) bending. Filled circles in (b) and (c) show the data taken when the sensor is flat and released. Solid lines are guide for the eye.

($\Delta(\rho_\parallel - \rho_\perp)$) must be zero. Thus, while the PHE voltage is unchanged, the AMR voltage must be affected by stress or strain due to the remaining $\rho_\perp$ in the Eq. 1. The data given in Figs. 3(a) and 4(a) supports this conclusion. Although the AMR voltage has been shifted to the higher values when the sensor shape has been changed from concave to convex, the change in the resistivity difference ($\Delta(\rho_\parallel-\rho_\perp)$) is negligible. In addition, the maximum and minimum

PHE voltages are almost constant under different bending conditions, indicating unchanged resistivity difference.

$$V_{AMR} = \frac{il}{wt}\left(\rho_\perp + (\rho_\| - \rho_\perp)\right)cos^2\theta \tag{1}$$

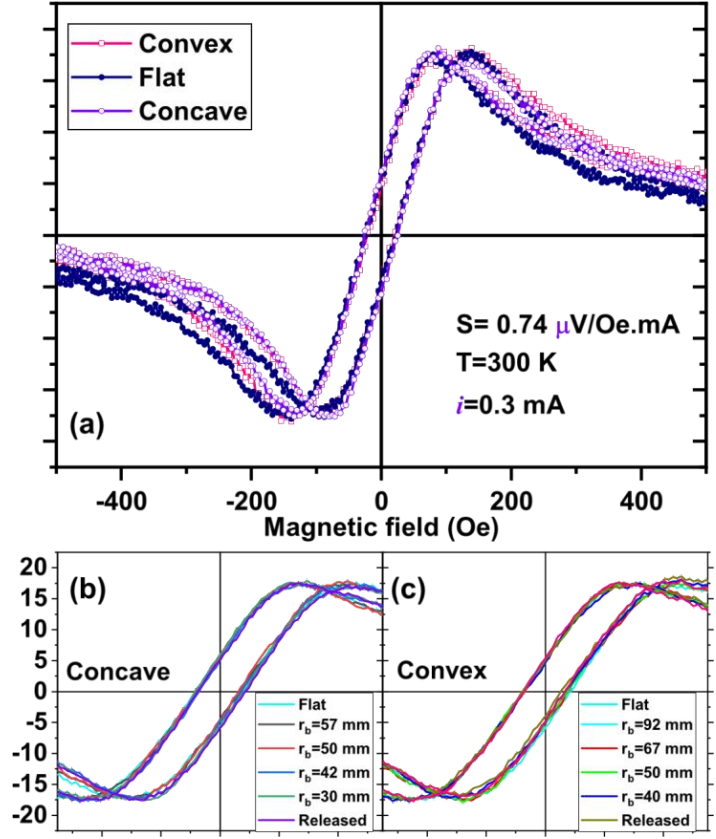

FIG. 4. (a) The PHE signals recorded at flat condition, as well as under maximum bending of the convex and concave configurations for bending radii of 40 mm and 30 mm, respectively. The PHE voltage profiles of the sensor as a function of bending radius for concave (b) and convex (c) conditions. Solid lines are guide for the eye.

$$V_{PHE} = \frac{i}{t}\left(\rho_\| - \rho_\perp\right)sin\theta cos\theta \tag{2}$$

Similarly, although the AMR ratio is inversely proportional to the $\rho_\perp$ (Eq. 3), the PHE sensitivity (Eq. 4) must be independent from the bending due to an unchanged PHE voltage. Namely, the PHE sensitivity in this work is not expected to be affected by a strain or stress produced by bending.

$$\text{AMR ratio} = \frac{\rho_\parallel - \rho_\perp}{\rho_\perp} \times 100 \qquad (3)$$

$$S_{PHE} = \frac{\Delta V_{PHE}}{\Delta H \cdot i} \qquad (4)$$

Furthermore, *Oh* and coworkers recently reported that the stress induced anisotropy is responsible for the observation of a change in the PHE sensitivity of a ring junction.[11] Indeed, in PHE sensors, the change in the magnetic anisotropies results in a shift of the peak positions along the magnetic field axis. However, this can be completely ruled out in the present work. Because the peak positions of the PHE signals presented in Fig. 4 do not reveal any changes in the magnetic field axis under concave and convex bending.

## IV.   CONCLUSIONS

In summary, we have successfully fabricated a shapeable planar Hall sensor with a sufficient magnetic field sensitivity for applications in human-machine interfaces. Real-time experiments indicate that the sensor states are very stable. In addition, the AMR voltages of this novel device are sensitive to the stress and strain produced by a concave and convex bending. However, we have reported that the simultaneously measured PHE signals remain unchanged under different bending conditions. These newly discovered features can lead to new application areas for the PHE sensors.

## ACKNOWLEDGMENTS

This work was supported by TÜBİTAK (The Scientific and Technological Research Council of Turkey) through the project number 116F083.